# TOPOLOGICAL DEFECTS AND HIGHEST ENERGY COSMIC AND GAMMA RAYS

G. SIGL

*Department of Astronomy & Astrophysics*
*Enrico Fermi Institute, The University of Chicago, Chicago, IL 60637-1433*

**Abstract.** In this paper we review the hypothesis that a substantial part of the cosmic ray flux observed above about $10^{19}$ eV may be produced by decaying or annihilating topological defects left over from phase transitions in the early universe at grand unification energy scales ($\approx 10^{16}$ GeV). Possible signatures of cosmic ray producing defect models are discussed which could be tested experimentally in the near future. We thereby focus on model independent universal spectral properties of the predicted particle fluxes.

## 1. Introduction

It is commonly believed that cosmic rays are mainly produced due to first order Fermi acceleration (see, e.g., Gaisser 1990) at astrophysical shocks. A potential source of cosmic rays (CR) of ultrahigh energies (UHE) (i.e. above $\approx 10^{18}$ eV) are relativistic shocks contained in radiogalaxies and active galactic nuclei (see, e.g., Cesarsky 1992; Rachen & Biermann 1993). The recent observation of cosmic rays above $10^{20}$ eV by the Fly's Eye (Bird, *et al.* 1993, 1994, 1995), and AGASA (Hayashida, *et al.* 1994; Yoshida, *et al.* 1995) experiments, and the experiment at Yakutsk (Efimov, *et al.* 1991; Egorov 1993) may, however, not be easily explained by this mechanism (Sommers 1993; Sigl, Schramm, & Bhattacharjee 1994; Elbert & Sommers 1995).

Therefore, it has been suggested that such superhigh energetic cosmic rays could have a non-acceleration origin (Hill 1983; Hill, Schramm, & Walker 1987; Bhattacharjee & Rana 1990; Bhattacharjee 1991; Bhattacharjee, Hill, & Schramm 1992; Sigl, Schramm, & Bhattacharjee 1994; Bhattacharjee & Sigl 1995) as, for example, the decay of supermassive elementary "X" particles associated with Grand Unified Theories (GUTs). These particles could be radiated from topological defects (TDs) formed in the early universe during phase transitions caused by spontaneous breaking of symmetries implemented in these GUTs [for a review on TDs, see Vilenkin (1985)]. This is because TDs, like ordinary and superconducting cosmic strings, domain walls and magnetic monopoles, are topologically stable but nevertheless can release part of their energy in the form of these X particles due to physical processes like collapse or annihilation. The X particles with typical GUT scale masses of the order of $10^{16}$ GeV decay subsequently into leptons and quarks. The strongly interacting quarks fragment into a jet of hadrons which results in typically of the order of $10^4$–$10^5$ mesons and baryons. It is assumed that these hadrons then give rise to a substantial fraction of the highest energy cosmic ray (HECR) flux as well as a considerable neutrino flux.

The shapes of the nucleon and $\gamma$-ray spectra predicted within such TD models are thus expected to be universal (i.e., independent of the specific process involving any specific kind of TD) and to be only dependent on the physics of X particle decay. This is because at HECR energies nucleons and $\gamma$-rays have attenuation lengths in the cosmic microwave background (CMB) which are small compared to the Hubble scale. Cosmological evolutionary effects which depend on the specific TD model are therefore negligible. In contrast, the predicted neutrino flux especially at lower energies would depend on the TD model.

Independently of the spectral shapes of the predicted nucleon and $\gamma$-ray fluxes the question arises whether a specific TD model is capable of producing a HECR flux at an observable level. In this paper we give an overview of the status of topological defect models for cosmic ray production with emphasis on spectral shapes and absolute flux levels. We organize the rest of the paper as follows: In section 2 we discuss the universal injection spectra of nucleons and $\gamma$-rays and evolve it to the spectra actually observed. Based on this we present in section 3 distinct signatures of these spectra which could be tested experimentally against other models of HECR production. The capability of various TD models of producing observable HECR fluxes is discussed in section 4.

## 2. The Universal Particle Spectra in Defect Models

In this section we discuss the universal nucleon and $\gamma$-ray spectral shapes within defect models. These are exclusively determined by the physics of X particle decay and the subsequent propagation of the decay products through intergalactic space.

### 2.1. Injection Spectra of Nucleons, $\gamma$-Rays and Neutrinos

We assume that each X-particle decays into a lepton and a quark each of an energy approximately half of the X particle mass $m_X$. The quark hadronizes by jet fragmentation and produces nucleons, $\gamma$-rays and neutrinos, the latter two from the decay of neutral and charged pions in the hadronic jets. The hadronic route is expected to produce the largest number of particles and we will focus on this channel.

The spectra of the hadrons in a jet produced by the quark are, in principle, given by Quantum Chromodynamics (QCD). Suitably parametrized QCD motivated hadronic spectra that fit well the data in collider experiments in the GeV–TeV energies have been suggested in the literature (Hill 1983). However, there is a great deal of uncertainty involved in extrapolating from the "low" energy data to the extremely high energies arising in the present situation. To study the sensitivity of results to the assumed hadronization spectrum, we will also use different injection spectra suggested by Chi, *et al.* (1993) on certain phenomenological grounds.

The injection spectrum, i.e., the number density of particles produced per unit time per unit energy interval, for the species a, where a can denote nucleons (N), $\gamma$-rays ($\gamma$) or neutrinos ($\nu$), can be written as

$$\Phi_\mathrm{a}(E_\mathrm{i}, t_\mathrm{i}) = \frac{dn_\mathrm{X}(t_\mathrm{i})}{dt_\mathrm{i}} \frac{2}{m_\mathrm{X}} \frac{dN_\mathrm{a}(x)}{dx}, \tag{1}$$

where $x \equiv 2E_\mathrm{i}/m_\mathrm{X}$, $E_\mathrm{i}$ being the energy at injection and $dN_\mathrm{a}/dx$ the effective fragmentation function describing the production of the particles of species a from the original quark. We will consider the following two cases for the fragmentation function:

(1) *QCD-motivated injection spectra*

In this case, the *total* hadronic fragmentation spectrum $dN_\mathrm{h}/dx$ is taken to be of the form (Hill 1983)

$$\frac{dN_\mathrm{h}(x)}{dx} = \begin{cases} \frac{15}{16} x^{-1.5}(1-x)^2 & \text{if } x_0 \leq x \leq 1 \\ 0 & \text{otherwise} \end{cases}, \tag{2}$$

where the lower cutoff $x_0$ is typically taken to correspond to a cut-off energy $\approx 1\,\mathrm{GeV}$. Assuming a nucleon content of $\approx 3\%$ and the rest pions, we can write the fragmentation spectra as (Bhattacharjee, Hill, & Schramm 1992; Aharonian, Bhattacharjee, & Schramm 1992)

$$\begin{aligned} \frac{dN_\mathrm{N}(x)}{dx} &= (0.03) \frac{dN_\mathrm{h}(x)}{dx}, \\ \frac{dN_\gamma(x)}{dx} &= \left(\frac{0.97}{3}\right) 2 \int_x^1 \frac{1}{x'} \frac{dN_\mathrm{h}(x')}{dx'} dx', \end{aligned} \tag{3}$$

where we omit the neutrino spectrum which is similar to the photon spectrum.

(2) *Phenomenological injection spectra*

Recently, somewhat different injection spectra have been suggested by Chi, *et al.* (1993) on the following phenomenological grounds. UHE $\gamma$-rays as well as protons generate lower energy $\gamma$-rays by $\gamma - \gamma_\mathrm{b}$ and $p - \gamma_\mathrm{b}$ collisions with background photons ($\gamma_\mathrm{b}$). The electromagnetic component of the energy lost in these collisions develops cascades in the universal radio background (URB), the cosmic microwave background (CMB) and in the infrared background (IRB) (in order of decreasing energy of the propagating photon). The measured flux of extragalactic $\gamma$-rays in the 100 MeV energy region (Fichtel, Simpson, & Thompson 1978) provides constraints on the form of the nucleon and gamma ray injection spectra at energies above $\approx 5 \times 10^{19}\,\mathrm{eV}$. Based on these considerations, Chi, *et al.* (1993) have suggested injection spectra which we describe by the following fragmentation functions:

$$\frac{dN_{\mathrm{N}}(x)}{dx} = A_{\mathrm{N}}\, x^{-1.5}\,, \quad \frac{dN_{\nu}(x)}{dx} = \frac{dN_{\gamma}(x)}{dx} = A_{\gamma}\, x^{-2.4}\,, \qquad (4)$$

with

$$\frac{A_{\gamma}}{A_{\mathrm{N}}} \approx 0.028 \left(\frac{10^{15}\,\mathrm{GeV}}{m_{\mathrm{X}}}\right)^{0.9}. \qquad (5)$$

The condition (5) comes from the requirement (Chi, *et al.* 1993) that the photon-to-nucleon ($\gamma/\mathrm{N}$) ratio at injection at energy $E = 10^{20}\,\mathrm{eV}$, i.e., at $x = 2 \times 10^{20}\,\mathrm{eV}/m_{\mathrm{X}}$ be $\approx 60$. The spectra (4) are absolutely normalized such that the total quark energy $m_{\mathrm{X}}/2$ is injected above $5 \times 10^{19}\,\mathrm{eV}$ (which we will use as our normalization point to the observed HECR flux)

2.2. THE EVOLVED SPECTRA

The evolution of the spectra is governed by energy-loss and/or absorption of the particles as they propagate through the extragalactic medium. Assuming uniform energy injection the diffuse flux (in units of number of particles per area, time, solid angle and energy) today ($t = t_0$), at energy $E_0$, is calculated by (see, e.g., Bhattacharjee & Rana 1990)

$$j(E_0) = \frac{3}{8\pi}\, ct_0 \int_0^{z_{\mathrm{i,max}}(E_0)} dz_{\mathrm{i}}(1 + z_{\mathrm{i}})^{-5.5} \frac{dE_{\mathrm{i}}(E_0, z_{\mathrm{i}})}{dE_0} \Phi(E_{\mathrm{i}}, z_{\mathrm{i}})\,. \qquad (6)$$

Here, $z_{\mathrm{i}}$ is the injection redshift corresponding to the injection time $t_{\mathrm{i}}$, $E_{\mathrm{i}}(E_0, z_{\mathrm{i}})$ is the necessary injection energy, and the maximum injection redshift $z_{\mathrm{i,max}}(E_0)$ is determined from the condition $E_{\mathrm{i}}(z_{\mathrm{i,max}}, E_0) \leq m_{\mathrm{X}}/2$.

In the continuous energy-loss approximation (Berezinsky & Grigor'eva 1988; Aharonian, Kanevsky, & Vardanian 1990) the general energy-loss equation in terms of redshift $z$ can be written as

$$\frac{1}{E}\frac{dE}{dz} = \frac{1}{1+z} + \frac{(1+z)^{1/2}}{H_0}\beta[(1+z)E]\,. \qquad (7)$$

The first term is due to redshift of the (relativistic) particle energy and the second term describes losses due to interactions with the background medium in terms of the energy-loss rate $\beta(E) = -(dE/dt)/E$ ($H_0$ is the Hubble constant). For a particle observed at the earth with an energy $E_0$ the necessary injection energy $E_{\mathrm{i}}(z_{\mathrm{i}})$ at redshift $z_{\mathrm{i}}$ can be found by integrating Eq. (7) back from $z = 0$ to $z = z_{\mathrm{i}}$ with $E_{\mathrm{i}}(z = 0) = E_0$.

For nucleons one can use the expression for $\beta(E)$ derived in Berezinsky and Grigor'eva (1988). The important process here is the Greisen-Zatsepin-Kuz'min (GZK) (Greisen 1966; Zatsepin & Kuz'min 1966) effect in which UHE nucleons above about $6 \times 10^{19}\,\text{eV}$ (today) lose energy drastically due to photopion production off the CMB photons. This gives rise to the onset of a sharp fall ("cutoff") of the extragalactic UHE nucleon flux. We neglect the secondary $\gamma$-ray and neutrino contribution caused by the interactions of UHE nucleons with the CMB.

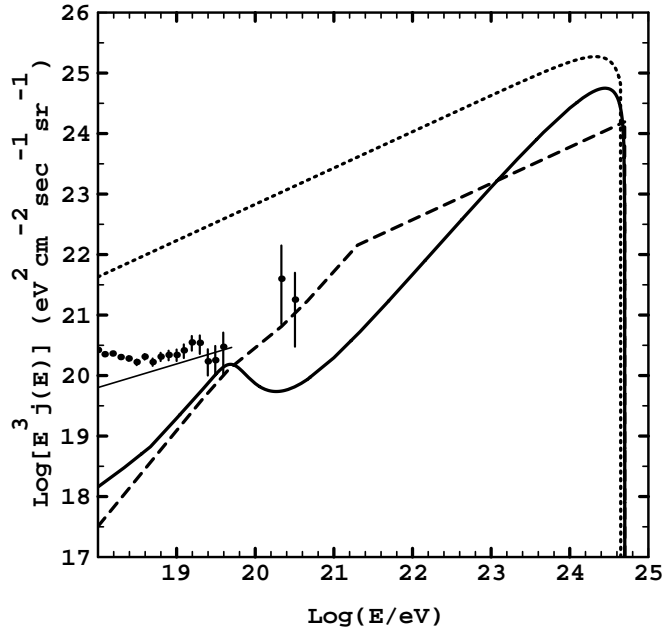

Fig. 1. Evolved neutrino (dotted line), $\gamma$-ray (dashed line) and proton (solid line) spectra typically predicted in defect models. An X particle mass of $m_X = 10^{16}\,\text{GeV}$ and the phenomenologically motivated injection spectra discussed in section 2.1 were used. The combined proton and gamma ray flux was normalized at $5 \times 10^{19}\,\text{eV}$ to the light (supposedly extragalactic) flux component (thin solid line) fitted (Bird, et al. 1993, 1994) to the high energy Fly's Eye data (dots with error bars, except the second highest energy event which was measured by the AGASA experiment). The two highest energy events observed would thus very likely have been $\gamma$-rays within the TD scenario.

The evolution of the UHE $\gamma$-ray spectrum is mainly governed by *absorption* of the UHE photons through $e^+e^-$ pair production on the CMB and on the URB. The developing electromagnetic cascades cause an increase of the effective penetration length of the UHE $\gamma$-rays. This effect, however, depends rather strongly on the poorly known strength of the intergalactic magnetic field. We therefore consider here only the absorption of the UHE $\gamma$-rays on the CMB and URB photons and use the absorption lengths given in Aharonian, Bhattacharjee, & Schramm (1992).

Similarly, UHE neutrinos can be absorbed by the thermal neutrino background but we are not so much interested in the neutrino fluxes here.

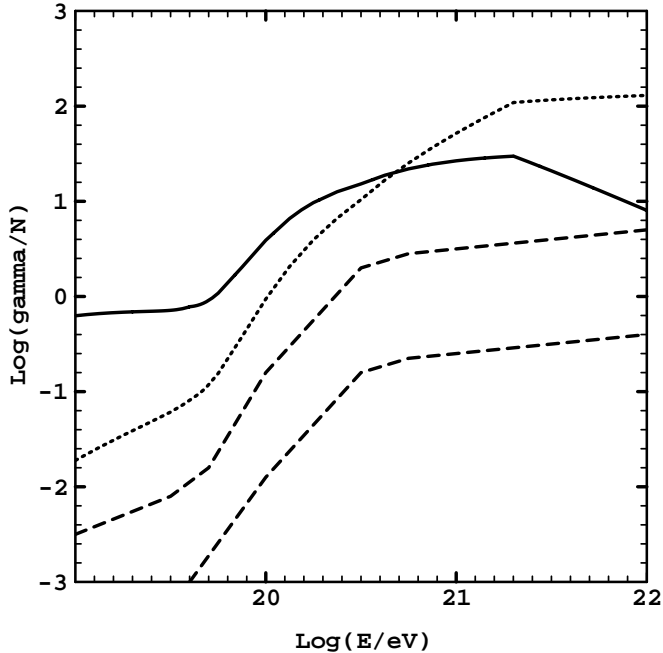

Fig. 2. The $\gamma/N$ ratios predicted within the QCD- and the phenomenologically motivated injection scenario in TD models (dotted and solid line, respectively) compared to the ones resulting from acceleration spectra with injection slopes 2 (upper dashed) and 3 (lower dashed).

Figure 1 shows an example for the spectra typically resulting. Only the neutrino spectrum depends on the cosmological history of X particle injection for which we assumed a $t^{-3}$ dependence which is typical for the more interesting TD scenarios (see section 4 below). Note that TDs alone are not able to explain the CR flux below a few $10^{19}$ eV without overproducing CRs at higher energies. More detailed calculations would have to solve the full Boltzmann equations for propagation and take non-uniform injection (discrete sources) into account. The latter effect can give rise to anisotropies in the HECR flux.

## 3. Signatures for Defect Sources of Highest Energy Cosmic Rays

As can be seen from Figure 1 one expects a $\gamma/N$ flux ratio which steeply increases with increasing energy around $10^{20}$ eV. This is mainly due to the fact that the nucleon attenuation length drops drastically above about $6 \times 10^{19}$ eV because of the GZK effect whereas the $\gamma$-ray attenuation length is slowly rising within this same energy range. A $\gamma/N$ ratio strongly increasing with energy is therefore expected not only within TD models. There is, however, a crucial difference between these and standard acceleration models which was first pointed out by Aharonian, Bhattacharjee, & Schramm (1992): Within TD models $\gamma$-rays dominate over nucleons already at injection whereas in case of shock acceleration $\gamma$-rays are only produced as secondaries of protons which are the primary accelerated particles. TD scenarios therefore predict a considerably higher $\gamma/N$ ratio. This is demonstrated in Figure 2 where the two injection scenarios from section 2.1 are compared with two acceleration scenarios which yield particularly high $\gamma/N$ values. With future instruments like the giant air shower array (Cronin 1995) it should be possible to distinguish $\gamma$-rays from nucleons at least on a statistical basis.

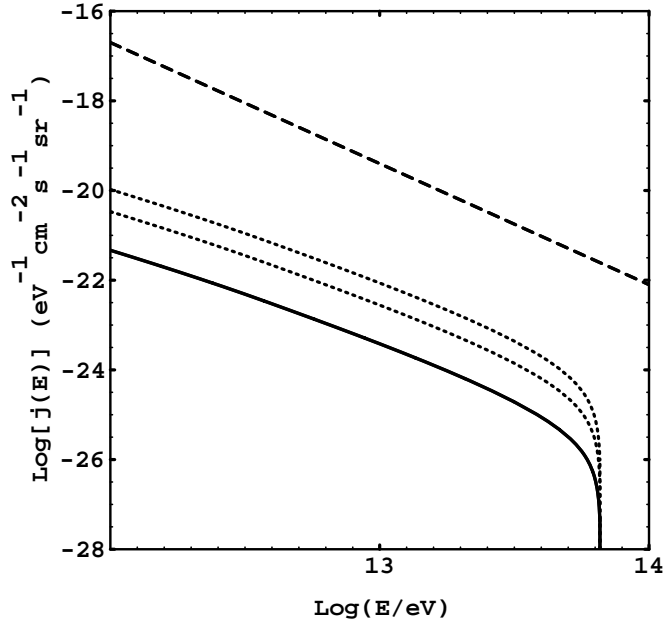

Fig. 3. Analytical approximation to the $\gamma$-ray fluxes below 100 TeV resulting from cascading in the CMB of $\gamma$-rays injected within defect models. The solid curve corresponds to the phenomenologically motivated injection scenario (independent of $m_X$). The upper and lower dotted curves are for QCD motivated injection for $m_X = 10^{16}$ GeV and $10^{15}$ GeV, respectively. Also shown is the approximate charged cosmic ray flux (dashed curve).

The high amount of high energy $\gamma$-ray injection in TD models leads to an enhanced electromagnetic cascade in the various photon backgrounds as compared to acceleration models. The injected energy is recycled below a certain threshold energy which today is of the order of 100 TeV. For cascading in the CMB alone the resulting cascade spectrum can be approximated analytically (Svensson & Zdziarski 1990). The spectra predicted for the defect models are shown in Figure 3. These fluxes are at least two orders of magnitude larger than what would be expected within acceleration models (Wdowczyk & Wolfendale 1990) which makes them a potential signature for a TD origin of HECR.

The HEGRA collaboration (Karle, *et al.* 1995) derived an upper limit on the $\gamma$-ray to charged CR flux ratio of about 1% above a few times $10^{13}$ eV. As can be seen from Figure 3 the QCD-motivated injection scenario for $m_X = 10^{16}$ eV comes near to this limit whereas the other scenario is still considerably below. It should be noted that accounting for background fields different from the CMB tends to decrease the predicted $\gamma$-ray fluxes in the energy interval shown in Figure 3.

## 4. Absolute Flux Calculation in Defect Models

Up to now we have simply normalized the HECR flux produced by TDs to the observed flux. Recently there was a claim (Gill & Kibble 1994) that TDs might not be able at all to produce observable HECR fluxes. We therefore briefly review this question here.

### 4.1. Ordinary Cosmic Strings

A cosmic string network can be formed in the early universe if for example a global or gauge symmetry group U(1) is spontaneously broken. The network is expected to reach the so called scaling regime where the length scale describing its large scale structure increases proportional to cosmic time $t$. In this regime closed string loops are continuously chopped off the network at a rate per volume $\approx \beta/t^4$ and with a typical length $\alpha t$. Here, $\alpha$ and $\beta$ are dimensionless constants given by string simulations (Austin, Copeland, & Kibble 1993). It has been shown (Bhattacharjee 1989; Bhattacharjee & Rana 1990) that cosmic strings can measurably contribute to the HECR flux only if a fraction $f \gtrsim 10^{-3}$ of the total energy in these loops is dissipated in the form of X particles on a time scale much smaller than the evolution time scale of the strings. In that case the X particle injection history is given by

$$\frac{dn_X}{dt} = f\alpha\beta\mu m_X^{-1} t^{-3}, \tag{8}$$

where $\mu$ is the square of the symmetry breaking scale. Gill & Kibble (1994) argue that this criterion is unlikely to be fulfilled under reasonable circumstances. However, we feel that this criterion could be satisfied, for example, in the following two cases: First, a considerable fraction of the loops forms in lowest frequency modes which lead to loop collapse within one oscillation period (Kibble & Turok 1982). This could happen if gravitational backreaction effects smooth out higher frequency wiggles on the loop. Or high-harmonic loops self-intersect and decay into smaller and smaller loops, finally releasing their energy in relativistic particles (Siemens & Kibble 1994). Second, hybrid systems of TDs could be formed such as light domain walls bounded by cosmic string loops. The domain walls in these systems could then aid the complete collapse of the string loops which form the boundaries of these walls.

### 4.2. SUPERCONDUCTING COSMIC STRINGS

Superconducting cosmic strings (SCS) (Witten 1985) as potential HECR sources were first proposed by Ostriker, Thomson, & Witten (1986). This was further pursued by Hill, Schramm, & Walker (1987). The relevant process is the ejection of charge carriers (which are the X particles in this case) carrying the electric current on the string once this current reaches a certain saturation value. The absolute X particle injection rate is in principle given by Eq. (8) with $f$ replaced by $L_{\rm s}/(100G\mu t)$ where $G$ is the Newtonian constant and $L_{\rm s}$ is the saturation length of the string loop.

The SCS scenario has become less promising because of several problems: First, it is possible that the SCS loops may never achieve the saturation current at all (Davis & Shellard 1989). Second, if the X particles decay within the high magnetic field region around the string the produced HECR are expected to be quickly degraded in energy due to synchrotron losses and other processes (Berezinsky & Rubinstein 1989). Third, it can be shown (Sigl, *et al.* 1995) that SCS which would substantially contribute to the HECR flux today would likely have produced spectral CMB distortions as well as light element abundances by $^4$He photodisintegration which violate experimental bounds. This is mainly because the X particle injection rate is proportional to $t^{-4}$ rather than $t^{-3}$ as for ordinary cosmic strings.

### 4.3. MAGNETIC MONOPOLES

Magnetic monopoles come from one particular type of TD solutions in spontaneously broken non-abelian gauge theories which are allowed in essentially all GUT models. Their formation in the early universe is inevitable in most models and leads to the well known "monopole problem" (for a review, see Kolb & Turner 1990), since their abundance is severely constrained. Besides the cosmological bound which limits the energy density in monopoles, the Parker bound limits their mean galactic abundance from their interactions with the galactic magnetic field (for an updated Parker bound, see Adams *et al.* 1993).

Monopoles ($M$) can be formed in metastable bound states with anti-monopoles ($\bar{M}$) ("monopolonium") which spiral in and finally collapse resulting in $M$-$\bar{M}$ annihilation. Monopolonium formed around the period of nucleosynthesis can thus give rise to HECR today (Hill 1983). The HECR flux can in principle be calculated from the monopole abundance and the relative monopolonium abundance at formation, $\xi_{\rm f}$ (Bhattacharjee & Sigl 1995). It turns out that, for example, for a monopole energy density $\approx 10^{-2}$ of the critical density, a symmetry breaking scale $m_X \approx 10^{16}\,\text{GeV}$ (which satisfies the Parker bound) and $\xi_f \approx 10^{-8}$ an observable HECR flux could be produced.

In addition, this scenario would not violate bounds on light element abundances and CMB distortions (Sigl, *et al.* 1995). Since the average distance between monopoles in this scenario is much smaller than average defect separations in other TD models one would expect comparatively isotropic fluxes.

In summary, SCS are highly unlikely to produce a substantial part of the observed HECR flux. In contrast, in our feeling ordinary cosmic strings and other TDs cannot be ruled out yet as HECR sources due to an incomplete understanding of the relevant physical processes involved. In particular, HECR production by magnetic monopole annihilation still seems to be an interesting option.

## Acknowledgements

This work was supported, in part, by the DoE, NSF and NASA at the University of Chicago, by the DoE and by NASA through grant NAGW-2381 at Fermilab, and by the Alexander-von-Humboldt Foundation.